\documentclass[
preprint,
 amssymb, amsmath,%
 aip,rsi,%
showpacs,
]{revtex4-1}

\usepackage{graphicx}
\usepackage[colorlinks=true,linkcolor=blue]{hyperref}

\begin{document}

\title{Neutron Resonance Spin Flippers: Static Coils Manufactured by Electrical Discharge Machining}

\author{N. Martin}
\affiliation{Heinz Maier-Leibnitz Zentrum (MLZ), Technische Universit\"at M\"unchen, Lichtenbergstr. 1, 85748 Garching, Germany}
\affiliation{Physik-Department E21, Technische Universit\"at M\"unchen, 85748 Garching, Germany}

\author{J.N. Wagner}
\affiliation{Karlsruhe Institute of Technology (KIT), Hermann-von-Helmholtz-Platz 1, 76344 Eggenstein-Leopoldshafen, Germany}
\affiliation{{\it work done at:} Forschungsneutronenquelle Heinz Maier-Leibnitz (FRM II), Technische Universit\"at M\"unchen, Lichtenbergstr. 1, 85747 Garching, Germany}

\author{M. Dogu}
\affiliation{Heinz Maier-Leibnitz Zentrum (MLZ), Technische Universit\"at M\"unchen, Lichtenbergstr. 1, 85748 Garching, Germany}

\author{C. Fuchs}
\affiliation{Heinz Maier-Leibnitz Zentrum (MLZ), Technische Universit\"at M\"unchen, Lichtenbergstr. 1, 85748 Garching, Germany}

\author{L. Kredler}
\affiliation{Heinz Maier-Leibnitz Zentrum (MLZ), Technische Universit\"at M\"unchen, Lichtenbergstr. 1, 85748 Garching, Germany}
\affiliation{Physik-Department E21, Technische Universit\"at M\"unchen, 85748 Garching, Germany}

\author{P. B\"oni}
\affiliation{Physik-Department E21, Technische Universit\"at M\"unchen, 85748 Garching, Germany}

\author{W. H\"au\ss ler}
\affiliation{Heinz Maier-Leibnitz Zentrum (MLZ), Technische Universit\"at M\"unchen, Lichtenbergstr. 1, 85748 Garching, Germany}
\affiliation{Physik-Department E21, Technische Universit\"at M\"unchen, 85748 Garching, Germany}

\date{\today}
\revised{\today}

\begin{abstract}
Radiofrequency spin flippers (RFSF) are key elements of Neutron Resonance Spin Echo (NRSE) spectrometers, which allow performing controlled manipulations of the beam polarization. We report on the design and test of a new type of RFSF which originality lies in the new manufacturing technique for the static coil. The largely automated procedure ensures reproducible construction as well as an excellent homogeneity of the neutron magnetic resonance condition over the coil volume. Two salient features of this concept are the large neutron window and the closure of the coil by a $\mu$-metal yoke which prevents field leakage outside of the coil volume. These properties are essential for working with large beams and enable new applications with coils tilted with respect to the beam axis such as neutron Larmor diffraction or the study of dispersive excitations by inelastic NRSE.
\end{abstract}


\maketitle
\section{Introduction}
\label{sec:intro}

Neutron Spin Echo (NSE) is a well-known scattering technique that allows exploring structural and magnetic dynamics in soft and condensed matter with sub-$\mu$eV energy resolution\cite{Mezei1972}. NSE methods are based on the controlled Larmor precession of the magnetic moment of the neutron in homogeneous magnetic fields. In the late 80's, R. G\"ahler and R. Golub have introduced the so-called Neutron Resonance Spin Echo \cite{Gaehler1987,Gaehler1992} (NRSE) technique by deriving the NSE principle on the basis of N. Ramsey's separated field technique \cite{Ramsey1950}. Following this conceptual breakthrough, several NRSE instruments have been developed in the last decade, opening up new perspectives in high resolution spectroscopy by drastically extending the accessible momentum (Q-) transfer range \cite{Keller1997,Koppe1999,Klimko2003,Haussler2004,Kawabata2006,Haeussler2007,Keller2007}. Similar to the NSE method, the NRSE energy resolution is -to first order- decoupled from the wavelength spread of the primary neutron beam. Thus, the technique can benefit from high primary neutron intensities, while achieving excellent energy resolution $10 \gtrsim \delta E \gtrsim 0.1~\mu$eV, as required for studies of slow motions with characteristic times in the 0.1-10 ns regime. 

NRSE is based on the Larmor precession of neutrons caused by static and high-frequency magnetic fields, produced by radiofrequency spin flippers (RFSF) which are installed in the spectrometer arms up- and downstream of the sample position. A RFSF consists of a $B_{\mathsf{0}}$ solenoid producing a static magnetic field and a radio-frequency coil responsible for the high-frequency magnetic field. The fine details of the dynamics in the sample are directly encoded in the beam polarization that is determined by an analyzer-detector unit. The spin manipulation device must consequently operate in a way that it preserves as much as possible the information contained in the neutron spin phase. Thus, the construction of coils delivering stable and homogenous magnetic fields over large volumes is one of the challenges in NRSE instrumentation. 

Several types of $B_{\mathsf{0}}$ coils have been built in order to comply with various instrument-specific requirements. For manufacturing, it appears that using round wires for the static coil facilitate the winding process. This method was used by various groups developing NRSE coils for thermal neutron beams\cite{Keller1993,Klimko2003a}. However, this technique is not compatible with long wavelengths ($\lambda \geq 8~\text{\AA}$) due to unavoidable randomization in the wire position leading to depolarization\cite{Prokudaylo2002}. Thus, on cold neutron based instruments\cite{Koppe1999,Bleuel2003,Haussler2004} or at pulsed sources\cite{Hayashida2007}, coils are normally wound using flat Al bands with larger cross sections. Such a design also offers a non-negligible advantage in terms of the maximum available current- and hence field- value. Indeed, when the wire diameter is too small, non tolerable resistive heating occurs. The Joule effect is, however, perfectly negligible in the case of flat bands with few mm thickness.

In this contribution we present the new concept of a $B_{\mathsf{0}}$ coil recently developed for the NRSE instrument RESEDA (Maier-Leibnitz Zentrum, Technische Universit\"at M\"unchen) that aims at combining the strengths of the otherwise used manufacturing techniques. In order to warrant a homogenous distribution of turns and a constant current density over the coil length the $B_{\mathsf{0}}$ body was produced by electrical discharge machining. The neutron window of the coil is large enough to realize a rotation of the coil of about 60$^{\circ}$ with respect to the through going neutron beam. This enables additional measurement options at RESEDA, e.g. wide-angle Larmor diffraction experiments\cite{Rekveldt2001}, besides improving the dynamical range for typical studies in the field of magnetism\cite{Kindervater2012} or soft-matter\cite{Marry2013}. We show first experimental tests of the new $B_{\mathsf{ 0}}$ coil, which was operated with a dedicated radio-frequency coil. In section \ref{sec:pi_flip}, we give a short introduction to theoretical elements underpinning the controlled neutron resonant $\pi$-flip. The new $B_{\mathsf{ 0}}$ coil design and properties are presented in section \ref{sec:coil_design}. We describe its experimental characterization in section \ref{sec:experimental} and deliver our conclusions and outlook in section \ref{sec:ccl}.

\section{\texorpdfstring{The resonant $\pi$-flip -- a short summary}{The resonant pi-flip - a short summary}}
\label{sec:pi_flip}

In NRSE, a pair of RFSFs separated by a field free region of length $L$ replaces the long solenoid used in the classical NSE technique. When two RFSFs produce subsequent $\pi$-flips of the beam polarization, the assembly simulates the Larmor precession of the neutron spin $\vec{\sigma}_{\mathsf{ n}}$ through an effective field integral $ 2 \times B_{\mathsf{0}} \cdot L$ where $B_{\mathsf{0}}$ is the static field produced inside the flippers. Technically, a RFSF creates a static magnetic field $\vec{B}_{0}$ encompassing a linearly oscillating field $\vec{B}_{\mathsf{1}}$. The latter can be described as the sum of two components $\vec{B}_{\mathsf{1}}^{\pm}$ of equal amplitude $\vec{B}_{1}$, counter-rotating with an angular frequency $\pm f_{\mathsf{rf}}$. The solution of the Schr\"odinger equation for the neutron spinor in such a field superposition  results in a time-dependent transition (or {\it spin flip}) probability modeled by the {\it Rabi equation}\cite{Rabi1937}:

\begin{equation}
	\mathcal{P}_{\pm\mp} = \frac{f_{\mathsf{1}}^{2}}{f_{\mathsf{Rabi}}^{2}} \cdot \sin^{2} \left( 2\pi f_{\mathsf{Rabi}} \cdot t_{\mathsf{f}} \right)
\label{eq:rabieq}
\end{equation}

where the {\it Rabi frequency} is given by

\begin{equation}
	f_{\mathsf{Rabi}} = \sqrt{\left( f_{\mathsf{rf}} - f_{\mathsf{0}} \right)^{2} + f_{\mathsf{1}}^{2}}
\end{equation}

with $f_{\mathsf{0}} = \gamma_{\mathsf{n}} B_{\mathsf{0}}$ and $f_{\mathsf{1}} = \gamma_{\mathsf{n}} B_{\mathsf{1}}$ the Larmor frequencies associated with the fields $B_{\mathsf{0}}$ and $B_{\mathsf{1}}$ respectively, through the neutron gyromagnetic ratio $\gamma_{\mathsf{n}} = 2.916~\text{kHz} \cdot \text{G}^{-1}$, and $t_{\mathsf{f}}$ is the time over which the oscillating field is applied. Because neutrons are moving in the field, $t$ is by definition the neutron time-of-flight through the RFSF, that is $t_{\mathsf{f}} = d/v$ with $d$ the coil thickness and $v = h / (m_{\mathsf{n}} \lambda)$ the neutron velocity, where $h$ is the Planck constant and $m_{\mathsf{n}}$ the neutron mass.

\begin{figure}[!ht]
	\includegraphics[width=16cm]{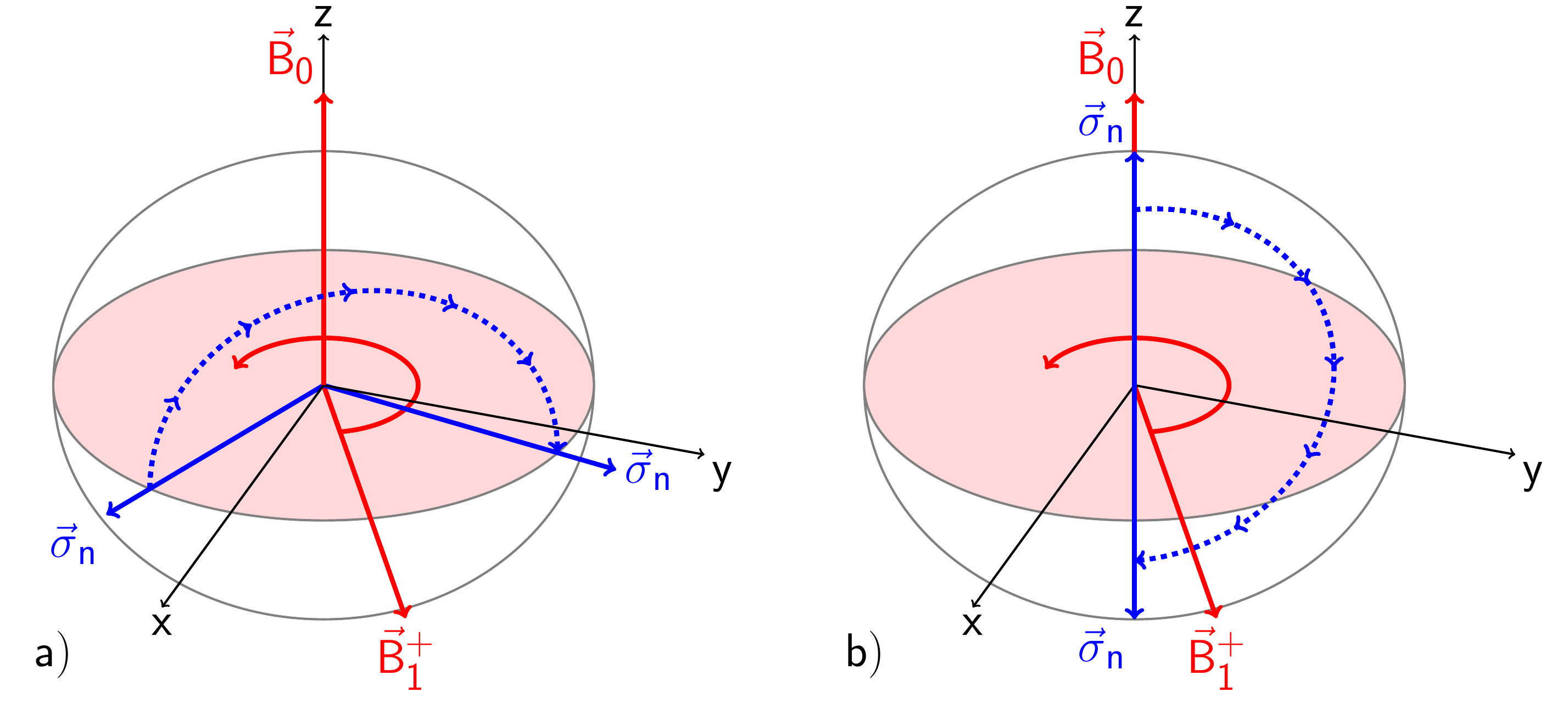} 
	\caption{Effect of the radiofrequency spin flipper on the neutron spin $\vec{\sigma}_{\mathsf{ n}}$. a) When the neutron is initially polarized in the $(x,y)$-plane and both resonance and $\pi$-flip conditions are satisfied, $\vec{\sigma}_{\mathsf{ n}}$ is rotated back to the horizontal plane after traversing the RFSF, mirrored with respect to $B_{\mathsf{1}}^{+}$. b) Using similar field values for the amplitudes of $\vec{B}_{\mathsf{0}}$ and $\vec{B}_{\mathsf{1}}^{+}$ but polarizing the neutron spin along the vertical $z$ direction ({\it i.e.} parallel to $\vec{B}_{\mathsf{0}}$), the effect of the RFSF is to flip its spin antiparallel to the quantization axis defined by $\vec{B}_{\mathsf{0}}$.}
	\label{fig:pi_flip}	
\end{figure}

When the {\it resonance condition} $f_{\mathsf{rf}} = f_{\mathsf{0}}$ is fulfilled, the probability $\mathcal{P}_{\pm\mp}$ oscillates in time between zero and unity while the Rabi frequency boils down to $f_{\mathsf{Rabi}} = f_{\mathsf{1}}$. Physically, it means that in the rotating coordinate system of $\vec{B}_{\mathsf{1}}^{+}$, the static field $\vec{B}_{\mathsf{0}}$ vanishes and the neutron spin initiates a Larmor precession around $\vec{B}_{\mathsf{1}}^{+}$ only\footnote{Under the condition $f_{\mathsf{1}} \ll f_{\mathsf{0}}$, only the component $\vec{B}_{\mathsf{1}}^{+}$ that rotates in the same direction as the Larmor precession of the neutron spin in the static field $\vec{B}_{\mathsf{0}}$ is relevant to the resonant $\pi$-flip and $\vec{B}_{\mathsf{1}}^{-}$ can be neglected. This is the so-called {\it rotating wave approximation}.}. Now, if one sets the amplitude $B_{\mathsf{1}}^{+}$ according to 

\begin{equation}
	 B_{\mathsf{1}} = \frac{v}{2 \gamma_{\mathsf{n}} e} \quad \Leftrightarrow \quad f_{\mathsf{1}} \cdot t_{\mathsf{f}} = \frac{1}{2} \quad ,
\label{eq:pi_flip_condition}
\end{equation}

one obtains a perfect mirroring of the incoming neutron polarization with respect to $\vec{B}_{\mathsf{1}}$ ({\it $\pi$-flip condition}, figure \ref{fig:pi_flip} a). If the beam is for instance initially polarized along $\vec{B}_{\mathsf{0}}$, a transition from {\it up} $|\!\uparrow\rangle$ to {\it down} $|\!\downarrow\rangle$ is induced. With such a configuration, that will be used for the experimental characterization of our new RFSF, the flipping efficiency of the coils is tested (figure \ref{fig:pi_flip} b).

From equation \ref{eq:rabieq}, one can obtain an expression for the spin projection $\langle \sigma_{\mathsf{z}} \rangle$ in the coordinate system defined by the relevant quantization axis of the problem, {\it i.e.} the unit vector directed parallel to $\vec{B}_{\mathsf{0}}$ that we define to be along the $z$-direction:

\begin{equation}
	\langle \sigma_{\mathsf{ z}} \rangle = 1 - \epsilon_{\mathsf{max}} \cdot \underbrace{\left( \frac{f_{\mathsf{1}}^{2}}{\left( f_{\mathsf{rf}} - f_{\mathsf{0}} \right)^{2} + f_{\mathsf{1}}^{2}} \right)}_{\textstyle \mathcal{L}} \cdot \left( 1 - \cos \left[ \frac{m_{\mathsf{n}} f_{\mathsf{Rabi}} d \lambda}{\hbar} \right]\right)
\label{eq:sigmaz}
\end{equation}

where $\epsilon_{\mathsf{max}}$ is the maximum flipping efficiency and the Lorentzian function $\mathcal{L}$ depends solely on the field amplitudes and oscillating frequency. In Eq. \ref{eq:sigmaz}, $\lambda$ appears explicitely and, because we are in general dealing with polychomatic beams, the effect of a wavelength distribution has to be accounted for. For the sake of simplicity, it is described by means of a Gaussian distribution of FWHM $\delta \lambda$ centered around the average value $\lambda_{\mathsf{0}}$, yielding

\begin{eqnarray}
	\nonumber
	\langle \sigma_{\mathsf{ z}} \rangle &=& 1 - \epsilon_{\mathsf{max}} \cdot \mathcal{L} \cdot \left( 1 - \int \cos \left[ 2 \pi f_{\mathsf{Rabi}} \cdot t \right] \cdot \frac{\exp \left( - 4 \ln 2 \left[ \frac{\lambda - \lambda_{\mathsf{0}}}{\Delta \lambda} \right]^{2}\right)}{\frac{\Delta \lambda}{2}\sqrt{\frac{\pi}{\ln 2}}} \cdot d\lambda\right)\\
	&=& 1 - \epsilon_{\mathsf{max}} \cdot \mathcal{L} \cdot \left( 1 - \cos \left[ \frac{m_{\mathsf{n}} f_{\mathsf{Rabi}} d \lambda_{\mathsf{0}}}{\hbar} \right] \cdot \exp \left[ - \frac{m_{\mathsf{n}} f_{\mathsf{Rabi}} d \Delta \lambda}{4 \hbar \sqrt{\ln 2}}\right]^{2}\right)
\label{eq:sigmazdlam}
\end{eqnarray}

Inspecting equation \ref{eq:sigmazdlam}, it appears that it is possible to model the effect of static field inhomogeneities by expanding $\mathcal{L}$ in Taylor series up to second order in $\Delta f_{\mathsf{0}} = \gamma_{\mathsf{n}} \Delta B_{\mathsf{0}}$, as follows

\begin{eqnarray}
	\nonumber
	\mathcal{L} \left( f_{\mathsf{0}} , \Delta f_{\mathsf{0}} \right) &\sim& \sum_{n} \frac{1}{n!} \cdot \frac{\partial^{n} \mathcal{L}}{\partial f_{\mathsf{0}}^{n}} \cdot \Delta f_{\mathsf{0}}^{n}\\
	\sim \mathcal{L} &\cdot& \left( 1 + \frac{3 \left(f_{\mathsf{0}}-f_{\mathsf{rf}}\right)^{2}- f_{\mathsf{1}}^{2}}{\left( \left(f_{\mathsf{0}}-f_{\mathsf{rf}}\right)^{2}+f_{\mathsf{1}}^{2} \right)^{2}} \cdot \Delta f_{\mathsf{0}}^{2} \right) \quad ,
\label{eq:lorexpand}
\end{eqnarray}

where the linear term is canceled due symmetry. Eq. \ref{eq:lorexpand} contains the leading terms in $\Delta f_{\mathsf{0}}$. Treating the cosine term of equation \ref{eq:sigmazdlam} is not necessary because it yields only 4$^{\mathsf{th}}$ (and higher) order terms that can be safely neglected. From these considerations, we see that measuring $\langle \sigma_{\mathsf{ z}} \rangle$ as a function of the static field $B_{\mathsf{0}}$ is an appropriate test of the flipper quality. This approach will be used in section \ref{sec:experimental} for characterizing quantitatively the properties of the new coil concept\footnote{The model presented in section \ref{sec:pi_flip} is derived in a semi-classical approach and can adequately describe the functionning of a RFSF. However, a full quantum approach yields further interesting insight into the NRSE technique which are out of the scope of this paper. Interested readers can find a thorough description of the interaction between a plane polarized wave and time-dependent magnetic fields\cite{Arend2011} and the formal solution of the so-called {\it Kr\"uger problem} in the literature\cite{Golub1994,Ignatovich2003}}.
\section{Coil design}
\label{sec:coil_design}

\begin{figure}[!ht]
	\centering
	\includegraphics[width=16cm]{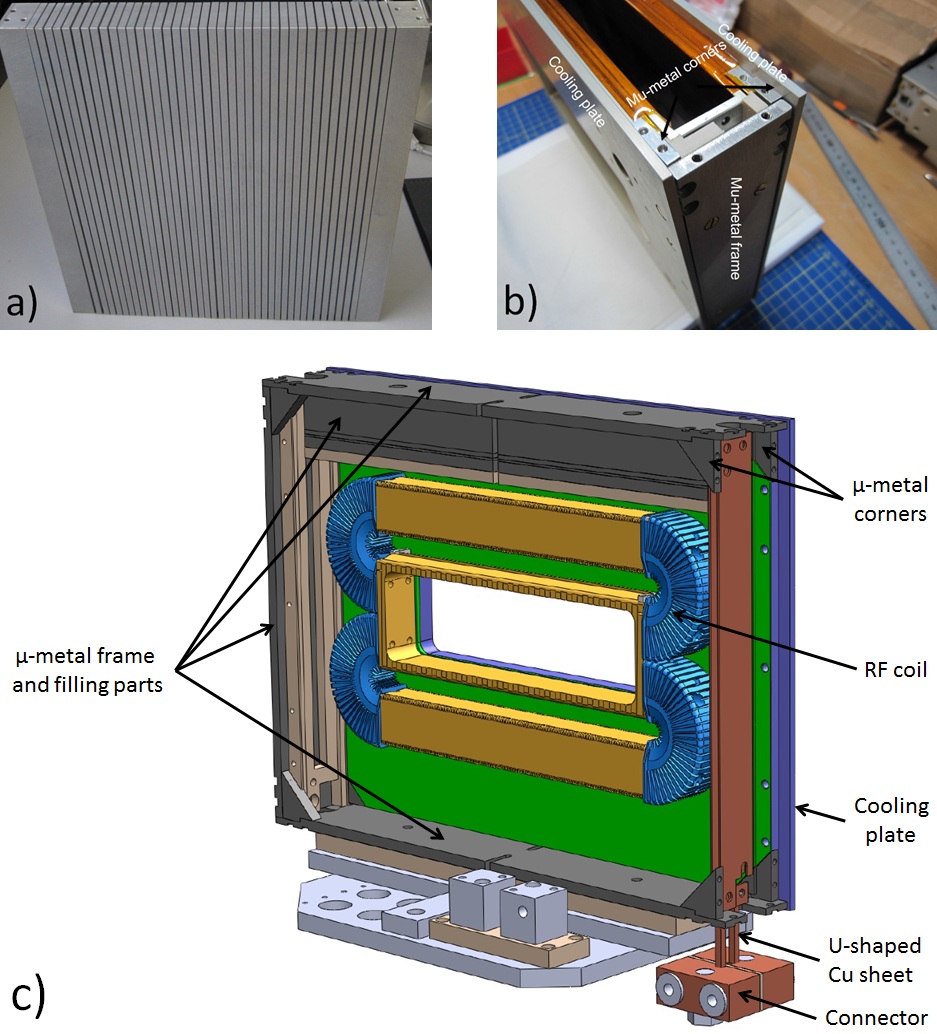}
	\caption{{\bf a)} -- Picture of the body of the newly designed $B_\mathsf{0}$ coil. {\bf b)} -- Sandwich design of the assembled coil. The $\mu$-metal frame, $\mu$-metal corners and cooling plates are depicted. {\bf c)} -- Technical drawing showing the main components of the coil design (see text). Note that the $\mu$-metal filling parts normally present at the bottom of the coil and identical to the one located at the top are omitted for clarity. Also shown is the small RF coil, inserted inside the $B_\mathsf{0}$ coil prior to operation.}
	\label{fig:b0_coil}	
\end{figure}

The body of the $B_0$ coil consists of a hollow, squared cylinder of AlMgSi with the dimensions $250 \times 197 \times 28$ mm$^3$ (h $\times$ w $\times$ d) and a wall thickness of 3\,mm. The windings are directly cut in the aluminum body by electrical discharge machining (EDM). This leads to a very homogenous height of 4mm for each windings and thus to a homogenous current density over the whole coil height. Forty-one horizontal cuts through 3 of 4 walls of the cylinder produce in total 40 windings over the whole height of the coil body. The winding process was actually realized by 40 tilted cuts which connect two adjacent horizontal cuts at one short body side (figure \ref{fig:b0_coil} a). The field which is produced by the non-horizontal winding process is compensated by a U-shaped sheet of copper which compensates the electrical field lines in the vertical direction. Additionally the copper conductor allows the simple connection of the coil to the power supply since both poles can be connected on the bottom of the coil. The $\mu$-metal frame is completed by a modular stack of $\mu$-metal sheets which are intended to fill in the empty space between RF coil placed in the center of the $B_\mathsf{0}$ coil and the $\mu$-metal frame. This will further homogenize the field produced by the coil. After cutting, the coil body was completely coated by para micron (C$_{12}$H$_{14}$Cl$_2$) using the chemical evaporation technique at 200$^\circ$C. This homogeneously distributed isolation layer between the windings has a thickness of only 15\,$\mu$m, which further reduces the inhomogeneities of the electrical stray field around the coil and provides a gap between two consecutive windings of only 30\,$\mu$m while isolating them perfectly against each other, due to its large electrical resistivity of $10~\text{M}\Omega\cdot\text{cm}$. The $\mu$-metal frame closely attached to the coil body guides and thus homogenizes the fields around the coil. By compression, it builds a perfectly flat surface as required. The corners of the frame are filled with small $\mu$-metal edges, to softly guide the field around the coil corners.
The default operation current is 100\,A, which roughly corresponds to a maximum field of about 270\,G ({\it i.e.} this corresponds to a resonance frequency of about 800\,kHz). To keep the coil body cold despite resistive heating, aluminum cooling plates are attached to the front and back surfaces. The furrow dug within each plate allows water circulation outside of the $70 \times 170$\,mm$^2$ (h $\times$ w) beam window at a nominal pressure of 4 bars. Heat transfer paste improves the thermal contact between coil and cooling plate. To ensure electrical insulation between the chiller system, the coil body and the $\mu$-metal frame, an anodic treatment was applied to the aluminum cooling plates. Estimating a neutron beam width of 30\,mm and a the total depth of the coil of 54\,mm (including cooling plates) a rotation angle of around 60$^{\circ}$ is possible without shielding the neutron beam. Note that this is only possible with RF coils which possess the same opening angle for neutrons. Figure \ref{fig:b0_coil} b shows one of two identical $B_\mathsf{0}$ coils which were produced as prototypes to test the newly applied coil concept. Figure \ref{fig:b0_coil} c further illustrates the design described above.

\section{Coil characterization}
\label{sec:experimental}

\begin{figure}[!ht]
	\centering
	\includegraphics[width=16cm]{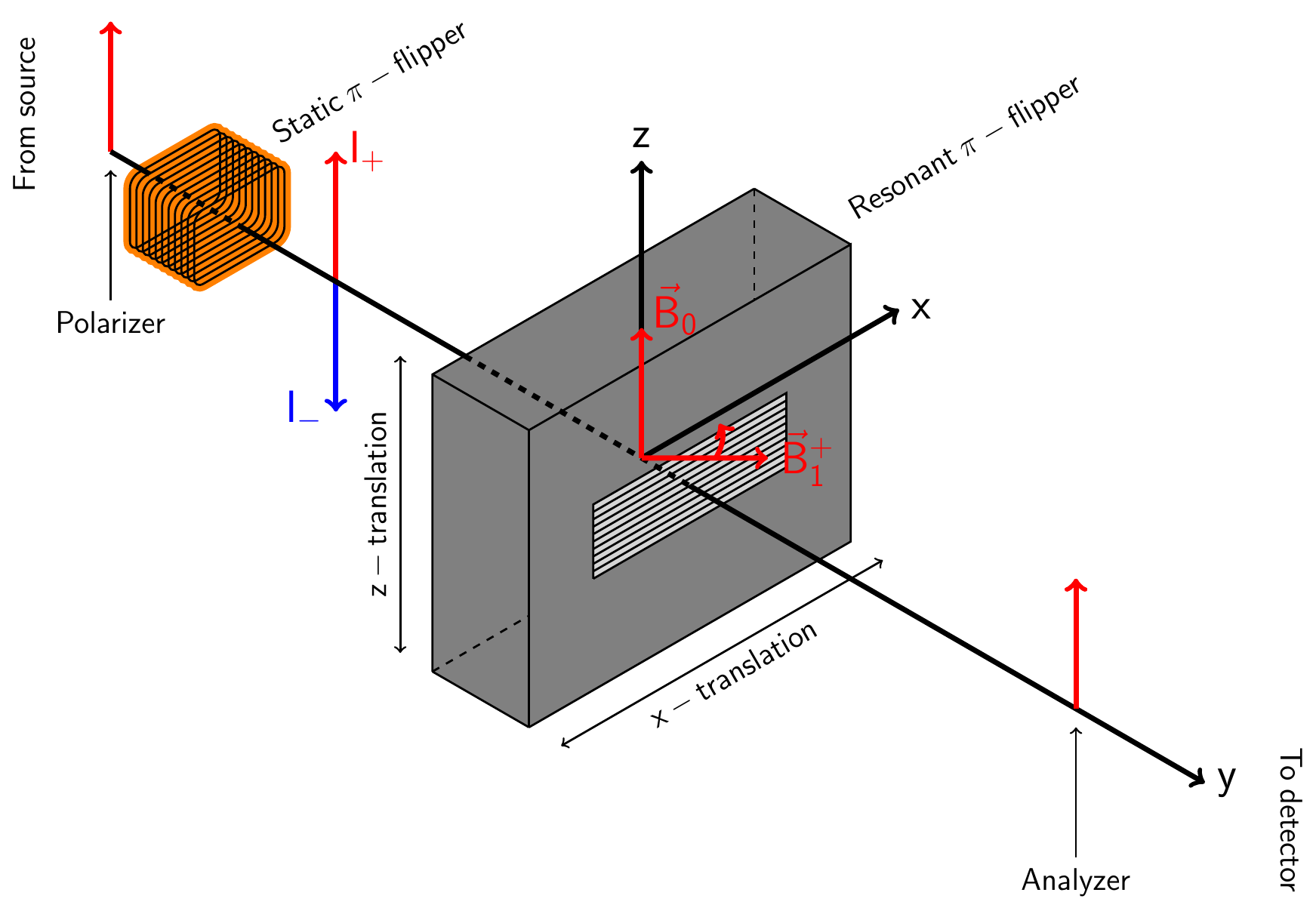}
	\caption{Experimental setup used for characterizing our RFSF. The whole setup was magnetically screened by a double $\mu$-metal shielding in order to avoid stray fields that could disturb the beam polarization.}
	\label{fig:exp_setup}
\end{figure}

The experiment we realized to test the performances of the RFSF was performed at the instrument RESEDA, where the beam polarization is prepared and analyzed by a pair of supermirror-based V-cavities\cite{Repper2012}. We have used a wavelength spectrum centered around $\lambda_{\mathsf{0}} = 5.42$ \AA~with a bandwidth of FWHM $\Delta \lambda / \lambda_{\mathsf{0}} = 0.18$ as provided by a velocity selector (Astrium NVS). The RFSF was installed in the double $\mu$-metal shielding surrounding the sample environment of the instrument and fixed on a two-dimensional translation table allowing for horizontal ($x$) and vertical ($z$) positioning with respect to the beam. At the coil position, the beam size was $10 \times 10$ mm$^{2}$. The experimental setup is schematically depicted in figure \ref{fig:exp_setup}. 
We have quantitatively examined the performances of the RFSF coil at four RF field frequencies over a fine mesh of $14 \times 4 = 56$ distinct positions in the $(x,z)$ plane. This corresponds to a total scanned area of $\Delta x \cdot \Delta z = 120 \times 30~\text{mm}^{2}$. These limits are set according to the dimensions of the RF coil placed within the $B_{\mathsf{0}}$ coil.

The $B_{\mathsf{0}}$ coil is fed by a static 150 V current supply (FUG NTN14000) whereas the RF coil is activated using a signal generator (HP Agilent 33120A) and a RF amplifier (T\&C Power Conversion AG 1016) supported by a variable capacity box. The RF current is dynamically regulated by reading back the voltage measured with a pick-up coil, which scales linearly with the RF field frequency.

At several $(x,z)$ positions, a short two- dimensional scan was performed by scanning the static and oscillating fields. Such a procedure allows for a systematic tracking of the optimal field configuration\cite{Haeussler1998}. This permits to experimentally fix the {\it $\pi$-flip condition} (equation \ref{eq:pi_flip_condition}) which is only a function of the neutron velocity, the latter being kept constant throughout the experiment.

In order to record the beam polarization as a function of $B_{\mathsf{0}}$, we have measured the beam intensity by activating ($I_{\mathsf{-}}$) and deactivating ($I_{\mathsf{+}}$) the static $\pi$-flipper located before the RFSF. This renders the beam polarization via

\begin{equation}
	\langle \sigma_{\mathsf{z}} \rangle = \frac{I_{\mathsf{+}} - I_{\mathsf{-}}}{I_{\mathsf{+}} + I_{\mathsf{-}}} \quad \cdot
\end{equation}

\begin{figure}[!ht]
	\centering
	\includegraphics[width=16cm]{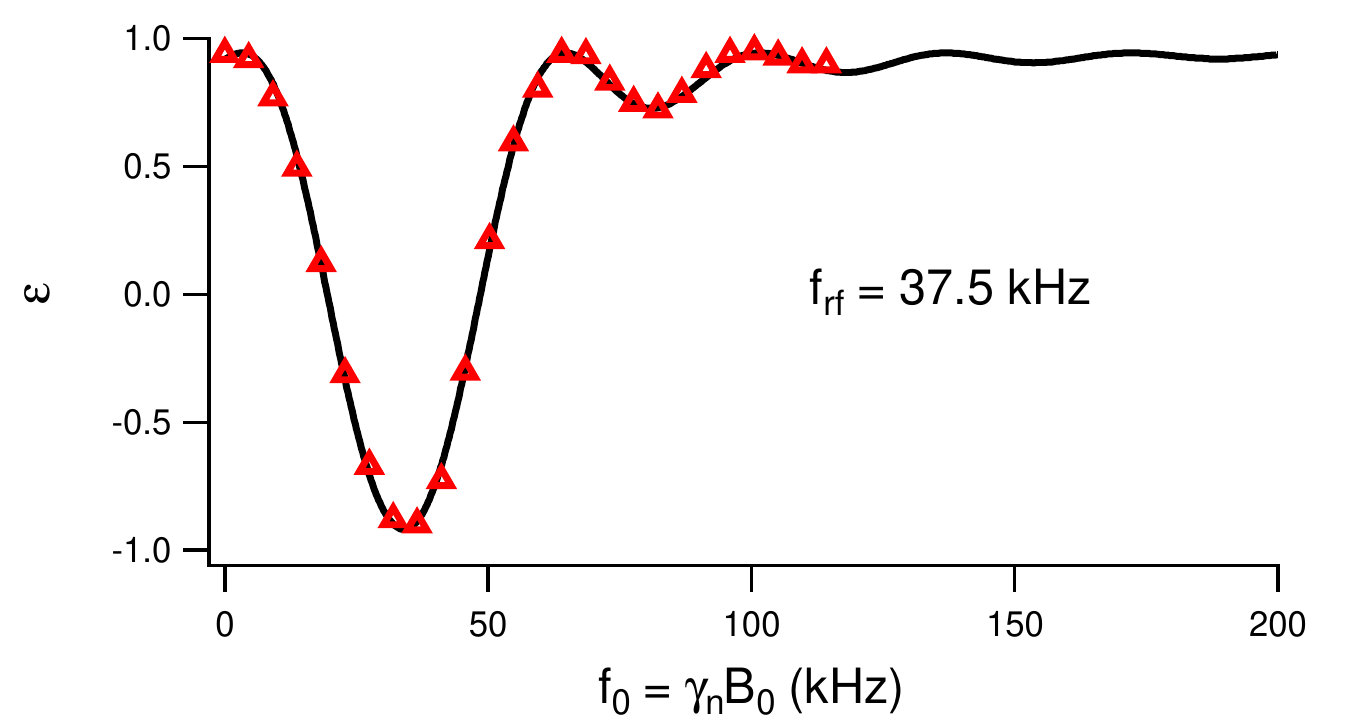}
	\caption{Typical $B_{\mathsf{0}}$-scan unveiling the magnetic resonance curve. Red triangles are experimental points and black line is a fit of equation \ref{eq:sigmazdlam} to the data (see text).}
	\label{fig:dataexample}
\end{figure}

The {\it flipping efficiency} is defined as $\epsilon = |\langle \sigma_{\mathsf{z}} \rangle| / P_{\mathsf{0}}$, where $P_{\mathsf{0}}$ is the polarization of the incident neutron beam, including the finite efficiency of the polarization analyzer. Figure \ref{fig:dataexample} shows a typical $\epsilon \left(f_{\mathsf{0}}\right)$ measurement. In order to extract all informations allowing to quantitatively characterize the $B_{\mathsf{0}}$-coil, the model equation \ref{eq:sigmazdlam} under the approximation \ref{eq:lorexpand} is fitted to the resonance curves measured at all points across the surface.

\begin{table}[!ht]
\centering
	\begin{tabular}{lccc}
	$f_{\mathsf{rf}}$ (kHz) & $\epsilon_{\mathsf{max}}$ & $\left. \Delta f_{\mathsf{0}} / f_{\mathsf{0}} \right|_{\mathsf{long.}}$  & $\left. \Delta f_{\mathsf{0}} / f_{\mathsf{0}} \right|_{\mathsf{trans.}}$\\
	\hline
	\hline
	37.5 & 0.94(1) & $3.0 \cdot 10^{-2}$ & $1.1 \cdot 10^{-2}$\\
	67.8 & 0.98(2) & $1.2 \cdot 10^{-2}$ & $5.5 \cdot 10^{-3}$\\
	132.4 & 0.97(2) & $2.6 \cdot 10^{-3}$ & $2.4 \cdot 10^{-3}$\\
	168.1 & 0.97(1) & $1.0 \cdot 10^{-3}$ & $1.8 \cdot 10^{-3}$\\
	\end{tabular}
	\caption{Quantitative outcome of the experimental characterization of the new RFSF. For each RF field frequency $f_{\mathsf{rf}}$, we display the measured values of the optimal flipping efficiency $\epsilon_{\mathsf{max}}$ (the standard deviation is given within brackets) and field homogeneities along $\left. \Delta f_{\mathsf{0}} / f_{\mathsf{0}} \right|_{\mathsf{long.}}$ and transverse $\left. \Delta f_{\mathsf{0}} / f_{\mathsf{0}} \right|_{\mathsf{trans.}}$ to the neutron beam.}
	\label{tab:results}
\end{table}

\begin{figure}[!ht]
	\centering
	\includegraphics[width=16cm]{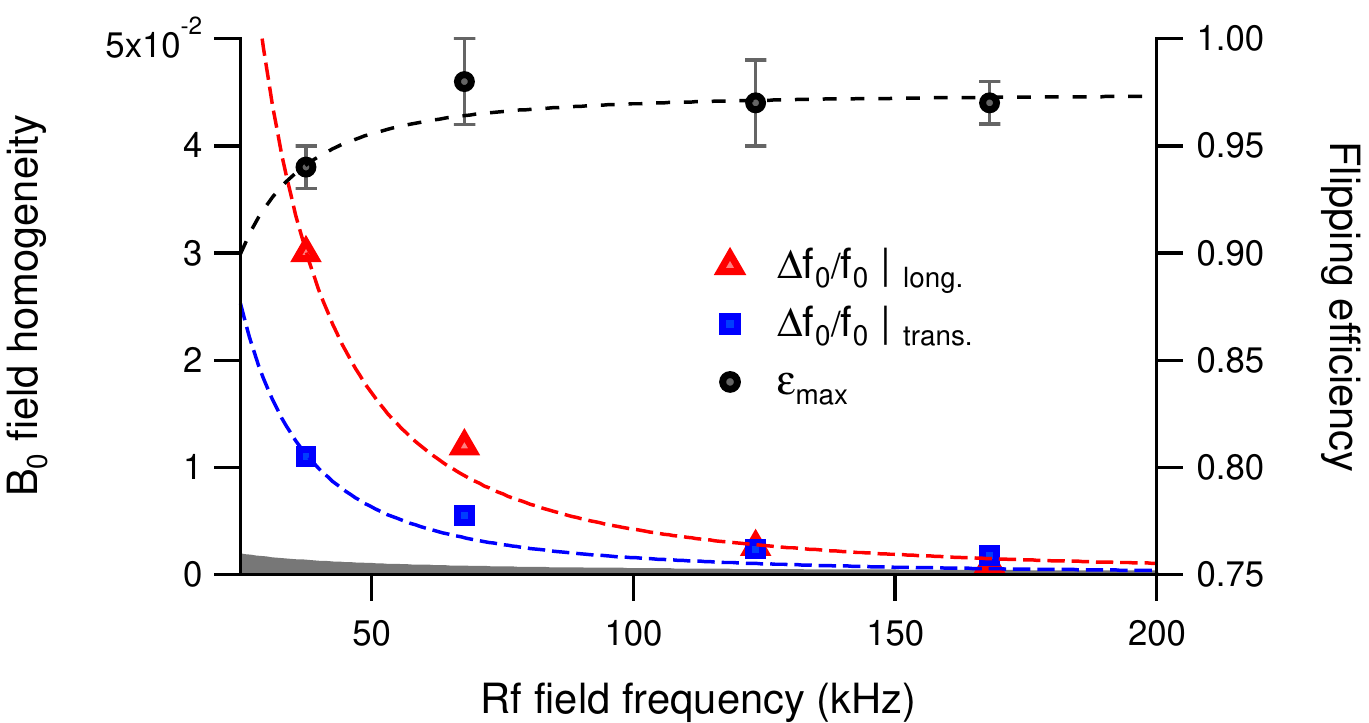}
	\caption{Evolution of the observed optimal flipping efficiency $\epsilon_{\mathsf{max}}$ and relative $B_{\mathsf{0}}$ field homogeneity as a function of the RF field frequency along (red triangles) and transverse (blue squares) to the beam axis. The dashed lines are fits of a model function $\propto f_{\mathsf{0}}^{-2}$ to the data, underscoring the trend imposed by the Siegert-Bloch shift. The gray area materializes the effect of a systematic error of the current positionning on the determination of $\left. \Delta f_{\mathsf{0}} / f_{\mathsf{0}} \right|_{\mathsf{long.}}$ and $\left. \Delta f_{\mathsf{0}} / f_{\mathsf{0}} \right|_{\mathsf{trans.}}$.}
	\label{fig:db0vsfreq}
\end{figure}

From the fitting procedure, we obtain the maximum flipping efficiency $\epsilon_{\mathsf{max}}$. In table \ref{tab:results}, we display the average value of $\epsilon_{\mathsf{max}}$ and its standard deviation over the coil surface. We observe that the polarization efficiency is almost constant over the full coil cross section. For $67.8 \leq f_{\mathsf{rf}} \leq 168.1~\text{kHz}$, the polarizing efficiency lies close to 1, the maximal achievable value under both resonance and $\pi$-flip conditions. This readily shows that the newly designed RFSF offer high performances. We note that at the lowest explored frequency ($f_{\mathsf{ rf}} = 37.5$\,kHz), a small drop of $\epsilon$ is evidenced. This is not linked with a misfunction of the coil but originates from the fact that we work with a linearly oscillating field. In the early stages of the development of the magnetic resonance technique, F. Bloch and A. Siegert\cite{Bloch1940} described a shift of the resonance frequency when a non-circular field is used and when the limit $f_{\mathsf{1}} \ll f_{\mathsf{0}}$ is violated. In the case of a RF neutron spin flipper, this translates into a drop of the flipping ratio a low $f_{\mathsf{ rf}}$ which evolves roughly as $1 - f_{\mathsf{0}}^{-2}$, as shown in figure \ref{fig:db0vsfreq}.

Additionally to the flipping efficiency, the homogeneity of $B_{\mathsf{0}}$ along the neutron path $\left. \Delta f_{\mathsf{0}} / f_{\mathsf{0}} \right|_{\mathsf{long.}}$ is derived from the fit. This parameter is actually the expansion variable $\Delta f_{\mathsf{0}}$ appearing in Eq. \ref{eq:lorexpand}. In table \ref{tab:results}, we give an account for the average static field deviation experienced by the throughgoing neutrons devided by the average value for $f_{\mathsf{0}}$. Similar to $\epsilon_{\mathsf{max}}$, its evolution as a function of the RF field frequency is entirely controlled by the Siegert-Bloch shift (figure \ref{fig:db0vsfreq}). 

Finally, the homogeneity of $B_{\mathsf{0}}$ across the coil surface can also be inferred from the position of the flipping maximum at fixed RF field frequency. This quantity is retrieved by letting the parameter $f_{\mathsf{rf}}$ in Eq. \ref{eq:sigmazdlam} float. As a figure of merit, we employ the observed normalized standard deviation $\left. \Delta f_{\mathsf{0}} / f_{\mathsf{0}} \right|_{\mathsf{trans.}}$ defined as 

\begin{equation}
	\left. \Delta f_{\mathsf{0}} / f_{\mathsf{0}} \right|_{\mathsf{trans.}} = \frac{\left( \frac{1}{n} \sum_{i=1}^{n} \left( f_{\mathsf{0}}^{i} - \overline{f_{\mathsf{0}}} \right)^{2} \right)^{\frac{1}{2}}}{\overline{f_{\mathsf{0}}}} \quad ,
\end{equation}

where $n = 56$ is the number of measured points in the $(x,z)$-plane, $f_{\mathsf{0}}^{i}$ the observed value at each point and $\overline{f_{\mathsf{0}}}$ the average value. The results for all RF field frequencies are gathered in table \ref{tab:results} and figure \ref{fig:db0vsfreq} again exemplifies the trend imposed by the Siegert-Bloch shift.

We note that the finite accuracy of the positioning of the current going through the $B_{\mathsf{0}}$-coil could hinder the proper determination of $\left. \Delta f_{\mathsf{0}} / f_{\mathsf{0}} \right|_{\mathsf{long.}}$ and $\left. \Delta f_{\mathsf{0}} / f_{\mathsf{0}} \right|_{\mathsf{trans.}}$ by artificially broadening the resonance curve. However, the stability of the power supply being of the order of $\Delta_{\mathsf{sys}} I \sim 5~\text{mA}$ (that is $\Delta_{\mathsf{sys}} f_{\mathsf{0}} \sim 40~\text{Hz}$), we find that this effect is in fact negligible (gray-shaded region in figure \ref{fig:db0vsfreq}). 

Taken together, the experimental results suggest that the coil works properly owing to the fact that the observed drop (increase) in flipping efficiency (field inhomogeneity) towards small RF field frequencies can be accounted for using very general arguments. 

\begin{figure}[!ht]
	\centering
	\includegraphics[width=16cm]{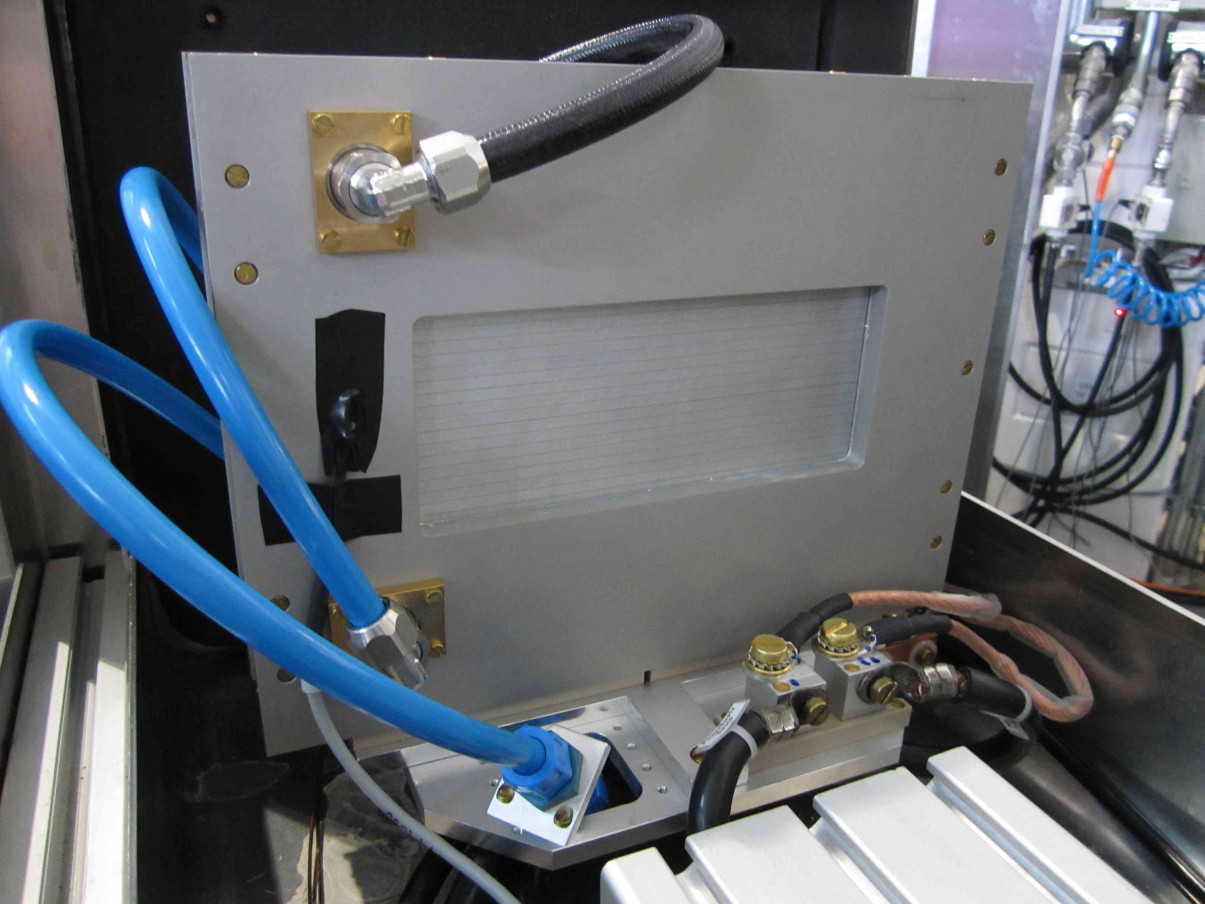}
	\caption{A new RFSF installed in the first arm of RESEDA. The electric and cooling water connections are visible.}
	\label{fig:coilRESEDA}
\end{figure}

During the course of a recent experiment, we have installed a pair of newly designed RF coils in the first arm of the instrument (figure \ref{fig:coilRESEDA}). The critical dynamics of Mn$_{1-x}$Fe$_{x}$Si alloys was studied with sub-$\mu$eV resolution\cite{Kindervater2014} by employing the MIEZE-SANS technique\cite{Brandl2011}. The neutron Larmor diffraction mode has also been tested by using the same setup. Taking $\lambda = 3$ \AA, which is the shortest wavelength accessible at the instrument, one obtains a maximum phase of $\varphi \simeq 3.6 \cdot 10^{4}$ and a theoretical resolution $\Delta d / d$ of the order of $2.8 \cdot 10^{-6}$. This value is at least one order of magnitude better than what can be obtained with conventional neutron diffractometers\cite{Ibberson2009,Torii2014}. The measurement of the thermal expansion curve of a pure Ge single crystal, with flippers tilted by 45$^{\circ}$ with respect to the beam axis, showed that this option works and the choice of the new coil design for an upgrade of the spectrometer meets the expectations.
\section{Summary and outlook}
\label{sec:ccl}
We showed that the new coil concept provides a homogenous and stable static magnetic field. The observed results lead to our conclusion that, even when the incidence angle of the coil becomes large with respect to the optical axis of the instrument, an excellent flipping efficiency shall be obtained. The homogeneity of this coil concept, which is widely based on the automated coil production process, favour the use of this advanced design for upgrading existing cold neutron based NRSE spectrometers or constructing new versatile instruments, for instance at the European Spallation Source. However, we insist that at any NRSE beamline, the actually available flipper efficiency can only be as high as the quality of the polarization transport through the whole setup. It means that efforts on the design of the magnetic screening of the neutron flight path throughout the instrument can not be avoided, as it is a very important parameter that eventually limits the achievable time (or energy) resolution. This is particularly true in the case of long NRSE/MIEZE spectrometers.
\section*{Acknowledgments}
\label{sec:ack}

We warmly thank R. Bierbaum, L. Sara\c{c} and M. Wipp (FRM II) for their involvement during the coil design process, as well as the preparation and course of the test experiments. This research project has been supported by the European Commission under the $7^{th}$ Framework Programme through the {\it "Research Infrastructures"}  action of the {\it "Capacities"} Programme, NMI3-II Grant number 283883.
\bibliographystyle{aipnum4-1}
\bibliography{b0paper}
\end{document}